\documentclass[%
 aip,
 amsmath,amssymb,
 reprint,%
]{revtex4-2}

\usepackage{graphicx}
\usepackage{dcolumn}
\usepackage{bm}

\usepackage[utf8]{inputenc}
\usepackage[T1]{fontenc}
\usepackage{mathptmx}
\usepackage{etoolbox}
\usepackage{amsmath}

\makeatletter
\def\@email#1#2{%
 \endgroup
 \patchcmd{\titleblock@produce}
  {\frontmatter@RRAPformat}
  {}{}
  {}{}
}%
\makeatother
\begin{document}

\preprint{AIP/123-QED}

\title[Physics of Fluids]{Microbial transport and dispersion in heterogeneous flows created by pillar arrays}
\author{Kejie Chen}
 \affiliation{School of Optoelectronic Engineering and Instrumentation Science, Dalian University of Technology, Dalian 116024, China}
\author{Kairong Qin}%
 \email{krqin@dlut.edu.cn}
\altaffiliation{ 
Corresponding author: Kairong Qin (krqin@dlut.edu.cn)
}%

\begin{abstract}
Swimming microbes, such as bacteria and algae, live in diverse habitats including soil, ocean and human body which are characterized by structural boundaries and heterogeneous fluid flows. Although much progress has been made in understanding the Brownian ratchet motions of microbes and their hydrodynamic interactions with the wall over the last decades, the complex interplay between the structural and fluidic environment and the self-propelling microbial motions still remains elusive. Here, we developed a Langevin model to simulate and investigate the transport and dispersion of microbes in periodic pillar arrays. By tracing the spatial-temporal evolution of microbial trajectories, we show that the no-slip pillar surface induces local fluid shear which redirects microbial movements. In the vicinity of pillars, looping trajectories and slowly moving speed lead to the transient accumulation and the sluggish transport of microbes. Comprehensive microscopic motions including the swinging, zigzag and adhesive motions are observed. In the pillar array of asymmetric pillar arrangements, the adjacent downstream pillars provide geometric guidance such that the microbial population has a deterministic shift perpendicular to the flow direction. Moreover, effects of the topology of the pillar array, fluid flowing properties and microbial properties on the microbial advection and dispersion in pillar arrays are quantitatively analyzed. These results highlight the importance of structures and flows on the microbial transport and distribution which should be carefully considered in the study of microbial processes.
\end{abstract}

\maketitle

\section{\label{Introduction}Introduction}
Microbes play a fundamental role in many environmental, medical and industrial processes, such as biofouling, biocorrosion and maintaining the homeostasis of human health \cite{1,2,3,4}. Microbial habitats, especially the surrounding fluid flows and the fractured spatial structures, strongly affect microbial activities and behaviors by modulating their transport, dispersion and distribution \cite{5, 6, 7, 8}. Unlike the passive particles which undergo convection and diffusion in fluid flows, microbes consume energy from nutrients and convert it into self-propelling motions \cite{9, 10, 11}. Consequently, the constant energy exchange between microbes and their environment drives the system out of equilibrium and generates complex dynamics and patterns. 

In the last decades, much progress has been made in understanding the microbial hydrodynamics by tracking microbial trajectories inside the artificial devices, such as the microfluidic channels \cite{12, 13, 14, 15, 16, 17, 17extra}. It has been found that, near the boundary, the heterogeneous flow velocity induces fluid shear on the microbe, leading to the frequent loops in the microbial trajectories. Due to the fluid shear, in a rectangular channel, a large portion of microbes are depleted from the low-shear regions and trapped in the high-shear regions \cite{12, 13, 14}. When the channel width decreases to a few microns, the stochasticity in the microbial orientation and their interactions with the wall cause microbes swimming against the fluid flows, the upstream swimming behavior \cite{15, 16}. In a curved channel, a large portion of microbes accumulate in the downstream of the channel corners \cite{17, 17extra}. In the presence of a circular pillar, microbes tend to reorient surrounding the pillar and attach to its leeward side \cite{17}. Moreover, in the natural habitats of microbes, such as the soils, solid grains produce spatial variability in fluid transport and change the distribution of passive colloids \cite{18, 19}. It is likely that the grain size and the heterogeneous flow velocity also influence the swimming microbes. However, until now, a comprehensive understanding of the interplay among flow, structure and microbial motility is still lacking. The rich class of the microscopic microbial dynamics and the underlying regulatory mechanisms of microbial activity by flows and structures need to be carefully studied.

Among the various artificial and natural structures, the array of cylindrical pillars is widely used in many industrial applications, such as the separation and filtration processes \cite{20, 21, 22, 23, 24}. When the medium flows through a symmetric pillar array, large-sized particles are captured between two pillars, while small particles pass the array by following the streamlines \cite{24}. For the swimming microbes whose sizes are smaller than the gap between two pillars, Dehkharguani et al. \cite{25} observed that microbes accumulate to form filamentous density patterns behind the pillars. And the densification patterns and the microbial dispersion coefficient depend upon the incident angle of the flow. In an asymmetric pillar array with deterministic lateral displacement (DLD), the asymmetric bifurcation of laminar flow around pillars cause a ‘zigzag mode’ of small particles and a ‘displacement mode’ of large particles, leading to a size-based displacement perpendicular to the flow direction \cite{20, 26, 27}. For the swimming microbes, Ranjan et al. \cite{28} have showed that the DLD pillar array can be used to separate rod-like bacteria, such as Escherichia coli (E.coli) and P.aeruginosa. But the microscopic dynamics of microbes in the DLD pillar array and the separation mechanisms have not been well understood, which hinders the design of the separation and filtration devices. Moreover, pillar arrays can also be regarded as a simplified and ideal representation of some porous media, such as the soils and structural tissues \cite{29}. Thus, understanding the microbial dynamics in both symmetric and asymmetric pillar arrays is also critical to study the microbial activity in the natural environments.

To investigate and quantify how flow gradients and self-propulsion determine the transport properties of microbes in complex structural environments, in this study we developed a Langevin model to simulate microbial activities in the pillar arrays. Specifically, the microbes are modelled as prolate ellipsoids with aspect ratio $q$, accounting for the hydrodynamic resistance of the microbial body and the flagellar bundles. The swimming speed $V$ of the microbe is assumed to be constant, and the swimming direction $\theta$ along the major body axis is influenced by the rotational Brownian motion or tumbling, hydrodynamic shear from the flow and the reflective structural boundaries. Based on the model, we showed that the no-slip surface of the pillar creates local velocity gradients, the fluid shear, and redirects microbial trajectories. Near the pillar surface, microbes move with a slower speed. Three types of motions, including swinging, zigzag and adhesive motions can be observed, especially when the pillar radius is large. Consequently, the advection of microbes in the flow direction is reduced, and a large number of microbes transiently accumulate near the pillar surface. The slow change of microbial density surrounding the pillar is reminiscent to the phenomenon when highly viscous fluid flows quickly through the center of the gap between two pillars but moves slowly near the pillar surface. In an asymmetric DLD pillar array, we found that the downstream pillars provide geometric guidance to the microbial movements, such that a large portion of microbes migrate perpendicular to the flow direction. The guiding effect of the neighboring pillars is also a macroscopic consequence of the microbes experiencing fluid shear and swimming toward the leeward side of the pillar. Moreover, in both symmetric and DLD pillar arrays, the horizontal and lateral microbial movements are influenced by the flow speed, the pillar size and arrangement, and the microbial properties. Overall, our work provides a quantitative model and a comprehensive understanding of the microbial dynamics controlled by the pillar geometries and fluid flows, which can provide helpful insights for a wide range of applications, from groundwater remediation, to microfiltration, to the design of active strategies and agents for drug delivery.

\begin{figure*}
\begin{center}
\includegraphics[width=2.0\columnwidth]{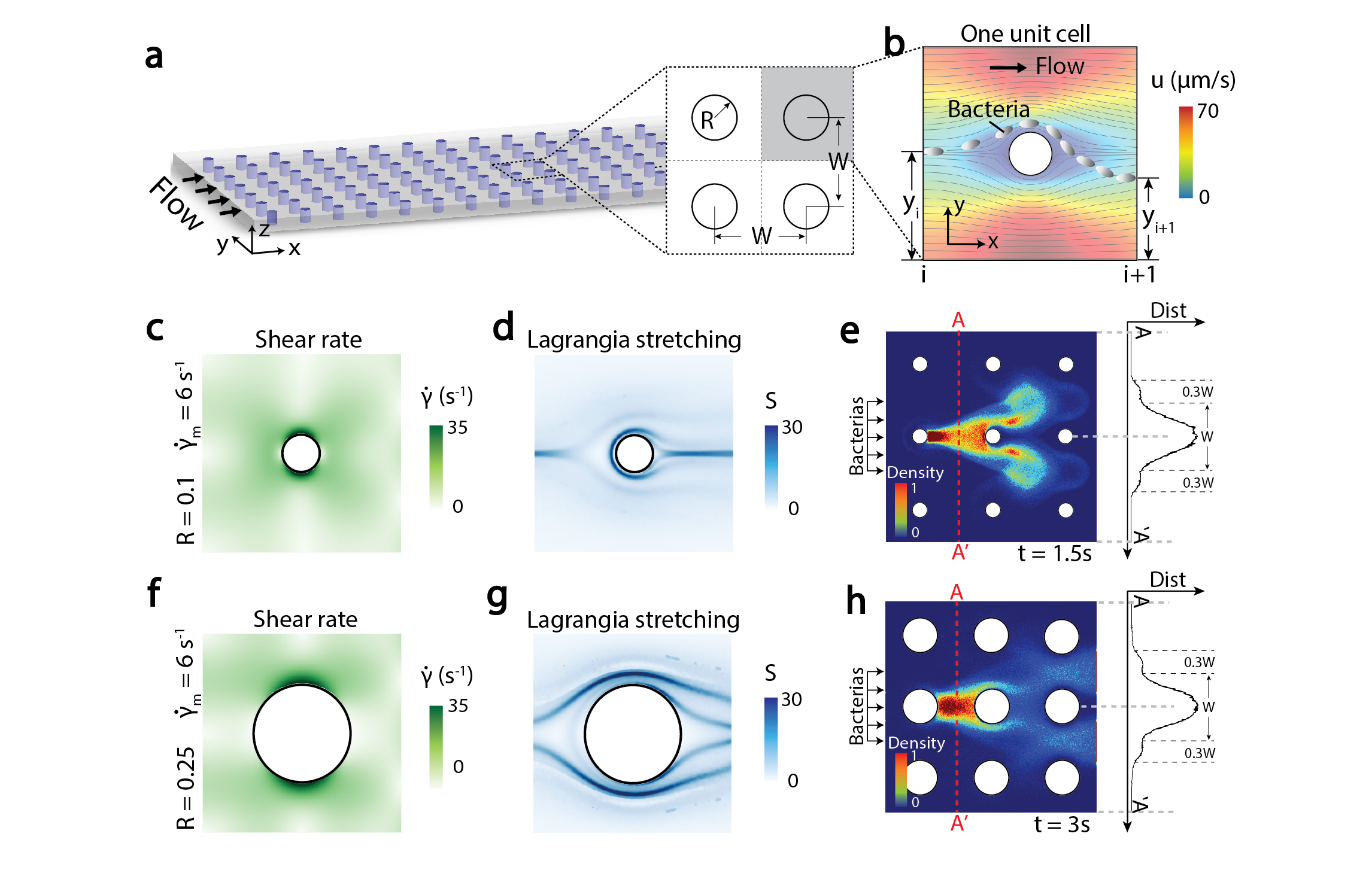}
\end{center}
\setlength{\abovecaptionskip}{-5pt}
\caption{\label{fig:wide}\textbf{The flow field and microbial distribution in an array of pillars.} \textbf{a,} Schematics of fluid flowing through a symmetric array of pillars. \textbf{b,} The distribution of flow speed in the horizontal direction ($x$ direction) $u$ in one repeated unit. \textbf{c,} The distribution of shear rate when the pillar radius $R$ is $0.1W$ ($W=100\, \mu m$) and the mean shear rate $\dot{\gamma}_m$ is 6 $s^{-1}$. \textbf{d,} The Lagrangia fluid stretching when $R=0.1W$ and  $\dot{\gamma}_m=6 \, s^{-1}$. \textbf{e,} The distribution of microbes at $t=1.5\, s$ in the 2D space, and the distribution of microbes that pass line AA’. \textbf{f,} The distribution of shear rate, \textbf{g,} the distribution of Lagrangia stretching, \textbf{h,} the distribution of microbes when $R=0.25W$ and $\dot{\gamma}_m=6\, s^{-1}$.}
\end{figure*}

\section{\label{Methods}Methods}
\subsection{The Langevin model of microbial motility in the flow}
We modeled a microbe as a prolate ellipsoid which has an effective aspect ratio $q$, a constant swimming speed $V$, and an effective rotational diffusion coefficient $D_r$. In the 2D space, the flow field at position $(x,y)$ is denoted as $\boldsymbol{U_f} (x,y)=[u(x,y),\, v(x,y)]$, where $u$ is the flow speed along $x$ direction and $v$ is the flow speed along $y$ direction. The microbe performs a run-and-tumble motion, determined by the flow speed $\boldsymbol{U_f}$, the swimming speed $V$, gradients of the flow field (i.e. $\dfrac{\partial u}{\partial x}$, $\dfrac{\partial u}{\partial y}$, $\dfrac{\partial v}{\partial x}$, $\dfrac{\partial v}{\partial y}$) and the rotational diffusivity $D_r$. The equations of microbial motion in the 2D flow are expressed as \cite{17}

\begin{align}
\dot{x} &= u + V cos \theta, \\
\dot{y} &= v + V sin \theta,
\end{align}
\vspace{-0.75cm}
\begin{multline}
\hspace{0.7cm}
\dot{\theta} = \dfrac{q^2-1}{q^2+1} \bigg[ \dfrac{1}{2} \bigg( \dfrac{\partial u}{\partial y} + \dfrac{\partial v}{\partial x} \bigg) cos2\theta  - \dfrac{\partial u}{\partial x} sin2\theta \bigg] \\
\hspace{1cm} - \dfrac{1}{2} \bigg( \dfrac{\partial u}{\partial y} - \dfrac{\partial v}{\partial x} \bigg) + \xi_r,
\end{multline}

Where $(x,y)$ is the microbial position. $\theta$ is the microbial moving direction, which is also the direction of microbial major body axis. $u$ is the background flow speed along $x$ direction. $v$ is the background flow speed along y direction. $V$ is the microbial swimming speed. $q$ is the aspect ratio. And $\xi_r$ is the rotational noise represented as a Gaussian-distributed angular velocity with mean zero and variance $2D_r/\Delta t$, $D_r$ is the rotational diffusion coefficient.

The equations of motion were solved using a fourth-order Runge-Kutta scheme implemented in Python. Trajectories of $10^7$ microbes are simulated. Based on the experimental measurements \cite{17}, for P. aeruginosa, the microbial parameters are set as $q=9.4$, $V=45 \, \mu m/s$ and $D_r=1.4 \, rad^2/s$. The time step $\Delta t$ is chosen as 0.001 $s$. The space discretization $\Delta x$ is chosen as 0.05 $\mu m$.

\subsection{Calculation of the flow field in the pillar array}
The background flow field $\boldsymbol{U_f}=[u,v]$ influenced by the size and arrangement of pillars is computed by numerically solving the 2D stokes equation, $\mu \nabla^2 \boldsymbol{U_f}- \nabla p=0$, in COMSOL Multiphysics, where $\boldsymbol{U_f}$ is the fluid velocity field, $p$ is the pressure and $\mu$ is the dynamic viscosity of water. Specifically, fluid flows through a rectangular chamber (500 $\mu m \times 1000 \, \mu m$) whose size is large enough such that the chamber boundaries have negligible effect on the microbial movements in the simulation time. Inside the chamber, there is an array of cylindrical pillars (5 columns, 10 rows). The pillar size is characterized by its radius $R$. Adjacent rows and columns of pillars are separated by a distance $W$ ($W=100 \, \mu m$) in both $x$ and $y$ directions, as shown in Figure 1a. In addition, in an asymmetric array of pillars, adjacent rows shift a distance $dH$ in the $y$ direction. Fluid flows with a constant speed from the left side of the chamber into the device. The right side of the chamber is set as a stress-free condition. And the two other sides of the chamber are set as periodic boundary conditions. And the pillar surfaces are set as the no-slip boundaries. The microbes flow into the pillar array from the center of the left side of the chamber where 200 $\mu m \leq y \leq 300 \, \mu m$.

\subsection{Quantification of the flow field properties}
The no-slip pillar surface creates local velocity gradients of the flow and influences microbial trajectories. To quantify the flow properties, distributions of shear rate and Lagrangian fluid stretching are calculated. The shear rate $\dot{\gamma}$, representing the velocity gradients, is calculated as the positive eigenvalues of the strain rate tensor $\boldsymbol{E}=\dfrac{1}{2}(\nabla \boldsymbol{U_f} + \nabla \boldsymbol{U_f}^T)$. 

The Lagrangian fluid stretching, which quantifies the deformation of a small circular fluid element into an ellipse due to the advection of ambient flow field, is calculated as the square root of the maximum eigenvalue of the left Gauchy-Green deformation tensors. Specifically, suppose a fluid particle which locates at position $\boldsymbol{x_0}$ at time $t=0$ is deformed by the flow field $\boldsymbol{U_f}$ and advected to a new position $\boldsymbol{\Phi}_{\lambda}(\boldsymbol{x_0})$ at time $t=\lambda$, where $\boldsymbol{\Phi}_{\lambda}=\boldsymbol{x_0}+\int_0^{\lambda} \boldsymbol{U_f} dt$ is the flow map. The Lagrangian fluid stretching is thus calculated as the eigenvalues of $\boldsymbol{C}_{\lambda}(\boldsymbol{x_0})=(\nabla \Phi_{\lambda}(\boldsymbol{x_0}))(\nabla \Phi_{\lambda} (\boldsymbol{x_0}))^T$. In a practical manner, we followed the approach of Parsa et al. \cite{30} to compute the Lagrangian stretching field. First, a regular grid of virtual particle position $\boldsymbol{x}$ ($500 \times 500$) within a unit cell of the pillar array (which contains one pillar) is chosen. For each virtual particle, we define four auxiliary points around it. The auxiliary points are located a small distance (0.01 $\mu m$) to the north, south, east and west from the virtual particle. By numerical integration of the flow field $\boldsymbol{U_f}$ backward in time (i.e. $-\lambda$), the trajectories of the virtual particle and the auxiliary points are obtained to get the flow map $\boldsymbol{\Phi_{\lambda}}$. During the numerical integration, every 0.01 $s$, the trajectories are rescaled to ensure the distances between auxiliary points and the virtual particle are always 0.01 $\mu m$. The flow map gradients, $\boldsymbol{\Phi}_{\lambda}(\boldsymbol{x_0})$, are then computed via the central differences using the four auxiliary points surrounding $\boldsymbol{x_0}$.

\subsection{Quantification of the microscopic microbial dynamics}
To quantify the microscopic microbial dynamics, the rotational Péclet number, defined as $Pe_r=\dfrac{<|\omega|>}{2D_r}$, is calculated, where $<|\omega|>$ is the mean absolute vorticity ($\omega=\nabla \times \boldsymbol{U_f}$), and $D_r$ is the rotational diffusion coefficient of the microbes. After obtaining the microbial trajectories by solving equations (1)-(3), the mean square displacement (MSD), the effective transverse velocity (the velocity perpendicular to the flow), the effective transverse dispersion coefficient, the effective lateral velocity (the velocity parallel to the flow) and the effective lateral dispersion coefficient are determined. The MSD is calculated as

\begin{equation}
MSD = \dfrac{1}{N} \sum_{i=1}^N [(x_i(t)-x_i(0))^2 + (y_i(t)-y_i(0))^2],
\end{equation}

where $N$ is the number of microbes ($N=10^7$ in this study), $(x_i (t),y_i (t))$ is the position of the microbe $i$ at time $t$. 

The maximum likelihood estimator (MLE) is used to infer the velocities and dispersion coefficients from microbial trajectories \cite{31, 32}. Specifically, suppose the microbial distribution along $x$ direction follows $p(x_i; \,v_x,\,D_x )=\dfrac{1}{\sqrt{4 \pi D_x \delta}} e^{- \frac{(x_i-v_x \delta)^2}{4D_x \delta}}$, where $x_i$ is the displacement of microbe $i$ along $x$ direction (i.e. $x_i=x_i (\delta)-x_i (0)$), $\delta$ is the time the microbe takes to achieve displacement $x_i$, $v_x$ is the lateral velocity, and $D_x$ is the lateral dispersion coefficient. The likelihood for obtaining $N$ microbial displacements $\{x_i\}$ is expressed as
\begin{equation}
f({x_i}; \, v_x, \, D_x) = \prod_{i=1}^N \dfrac{1}{\sqrt{4 \pi D_x \delta}} e^{- \frac{(x_i-v_x \delta)^2}{4 D_x \delta}},
\end{equation}
By solving $\arg \max f(x_i; \, v_x, \, D_x)$, we get the estimated lateral velocity $\hat{V}_x$ and the estimated lateral dispersion coefficient $\hat{D}_x$ as
\begin{align}
\hat{V}_x &= \dfrac{1}{N \delta} \sum_{i=1}^N x_i, \\
\hat{D}_x &= \dfrac{1}{2N \delta} \sum_{i=1}^N (x_i-\hat{V}_x \delta)^2,
\end{align}
Similarly, based on the microbial displacements in the $y$ direction ${y_i}$, the transverse velocity $\hat{V}_y$ and the transverse dispersion coefficient $\hat{D}_y$ can be estimated using the MLE as
\begin{align}
\hat{V}_y &= \dfrac{1}{N \delta} \sum_{i=1}^N y_i, \\
\hat{D}_y &= \dfrac{1}{2N \delta} \sum_{i=1}^N (y_i-\hat{V}_y \delta)^2.
\end{align}

\begin{figure*}
\begin{center}
\includegraphics[width=1.8\columnwidth]{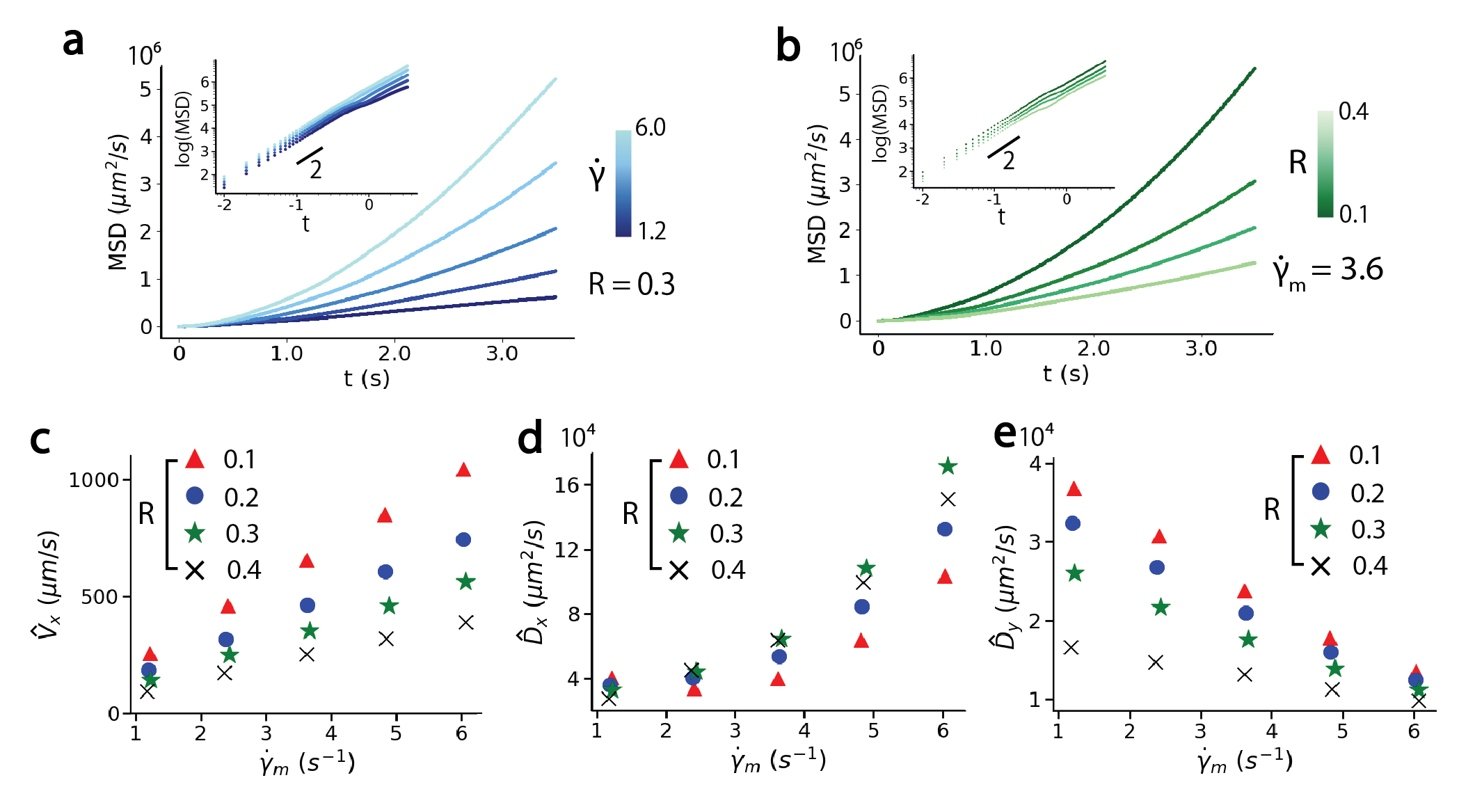}
\end{center}
\setlength{\abovecaptionskip}{-5pt}
\caption{\label{fig:wide}\textbf{The microbial motility influenced by the pillar radius $R$ and the mean fluid shear $\dot{\gamma}_m$ in a symmetric pillar array.} \textbf{a,} The MSD influenced by the mean fluid shear $\dot{\gamma}_m$. \textbf{b,} The MSD influenced by the pillar radius $R$. \textbf{c,} The estimated lateral microbial velocity $\hat{V}_x$, \textbf{d,} the estimated lateral diffusion coefficient $\hat{D}_x$, and \textbf{e,} the estimated transverse diffusion coefficient $\hat{D}_y$ over the mean fluid shear $\dot{\gamma}_m$ influenced by the pillar radius $R$.}
\end{figure*}

\section{Results}
\subsection{Microbes flow through a symmetric pillar array}
We investigated the effect of heterogeneous flow created by the pillar array on swimming microbes by tracking their trajectories and distributions. In a rectangular chamber consisting of a periodic array of circular pillars, fluid flows from the left side of the chamber to the right, and microbes enter the chamber from the center of the left inlet and swim with the flow (Figure 1a). The pillar radius $R$ varies from 10 $\mu m$ to 40 $\mu m$. The spacing between two adjacent pillars $W$, which is also the width and height of one periodic unit, is held constant ($W=100 \, \mu m$). The mean fluid flow speed varies from 0 to $400 \, \mu m/s$, corresponding to a mean shear rate $\dot{\gamma}_m$ approximately ranging from 0 to 10 $s^{-1}$. Figure 1b shows the flow streamlines in one unit. Due to the symmetry, the position a streamline enters the unit, denoted as $y_i$, is the same as the position it leaves the unit, denoted as $y_{i+1}$. Therefore, small-size passive particles, including the non-spherical particles, flow through the unit following the streamlines. And the particle distribution at the right outlet of the unit is the same as the distribution at the left inlet when diffusion is negligible. However, since the ellipsoidal microbes have active swimming ability, their trajectories are deviated from the flow streamlines ($y_{i+1} \neq y_i$). Specifically, the fluid shear $\dot{\gamma}$, as shown in Figure 1c and f, align the major axis of the microbial elongated body perpendicular to the pillar surface. In the vicinity of the pillar, the swimming speed of the microbe is larger than the local flow velocity, so the microbes will be able to swim upstream and reach the leeward side of the pillar. Moreover, to consider the nonlocal historical effects stemming from advection, we further calculated the Lagrangia fluid stretching field over a time interval $\lambda=3.5 \, s$ (Figure 1d and g). Since the microbial density is strongly correlated with the regions of high stretching \cite{25}, Figure 1d and g indicate that microbes that locate close to the pillar tend to be attracted to swim around the pillar. When the pillar radius increases, the influential distance within which the microbes might be captured by the pillar becomes larger. Figure 1e and h show that a large number of microbes accumulate behind the pillar due to the bending of microbial trajectories by the local fluid shear. And the time for most microbes arriving at the leeward side of the pillar is influenced by its radius $R$. Generally, when the pillar radius increases, the population of microbes move slower around the pillar, and the time microbes take to accumulate at the leeward pillar side is larger (Figure S1). However, by tracking the position where a microbe passes line AA’ (line AA’ locates between the first and second columns of pillars), we found that the pillar radius has only slight influence on the microbial distribution (Figure 1e and h, Figure S1). When $R=0.1W$ and $R=0.25W$, microbial distributions have a large peak at the center and heavy tails at the two sides (Figure 1e and h). In addition, microbes that locate far from the pillar pass line AA’ earlier than microbes near the pillar surface (Figure S1). In a macroscopic point of view, the reduction of the flowing speed of microbial population by the pillar geometries is reminiscent to the dynamics of highly viscous fluid flowing through the pillar array. Nevertheless, the fluid viscosity arises from the internal frictional force between adjacent layers of fluid. But the slower changes of microbial densification patterns are caused by the interactions between individual microbes and the pillar surface.

\begin{figure*}
\begin{center}
\includegraphics[width=1.8\columnwidth]{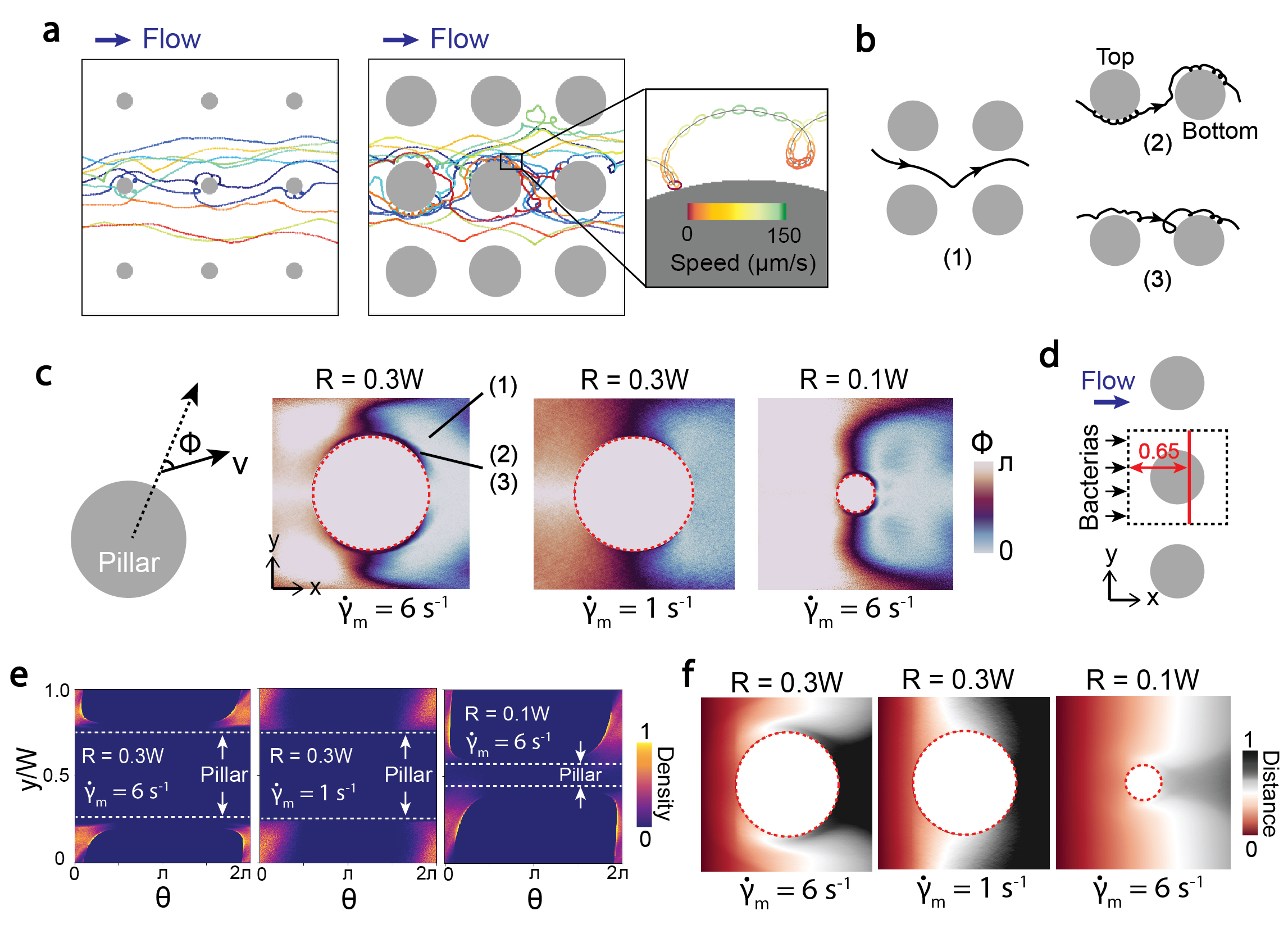}
\end{center}
\setlength{\abovecaptionskip}{-5pt}
\caption{\label{fig:wide} \textbf{The microscopic microbial dynamics surrounding a pillar.} \textbf{a,} Trajectories of microbes in pillar arrays (left $R=0.1W$, right $R=0.3W$) when the mean shear rate $\dot{\gamma}_m$ is 6 $s^{-1}$. Different trajectories are labeled in different colors. In the small rectangular window, the trajectory near the surface of the pillar is shown, where the color represents the microbial speed, and the ellipsoidal shape indicates the long body axis (also the moving direction $\theta$). \textbf{b,} Three types of microbial movements. Specifically, (1) ‘swinging’ mode, (2) ‘zigzag’ mode, and (3) ‘adhesive’ mode. \textbf{c,} Distributions of microbial moving direction $\phi$ influenced by the pillar radius $R$ and the mean shear rate $\dot{\gamma}_m$ in a circular coordinate where the coordinate center is the pillar center. (1), (2) and (3) in \textbf{c} refer to the three types of motion shown in \textbf{b}. \textbf{d,} Schematics showing the line (labelled in red) where the microbial densities are measured. The distance between the left inlet and the red line is $0.65W$. \textbf{e,} Distributions of microbial densities along the red line in \textbf{d} in phase space $(y-\theta)$ influenced by $R$ and $\dot{\gamma}_m$. \textbf{f,} Distributions of microbial moving distances influenced by $R$ and $\dot{\gamma}_m$.}
\end{figure*}

\begin{figure*}
\begin{center}
\includegraphics[width=1.8\columnwidth]{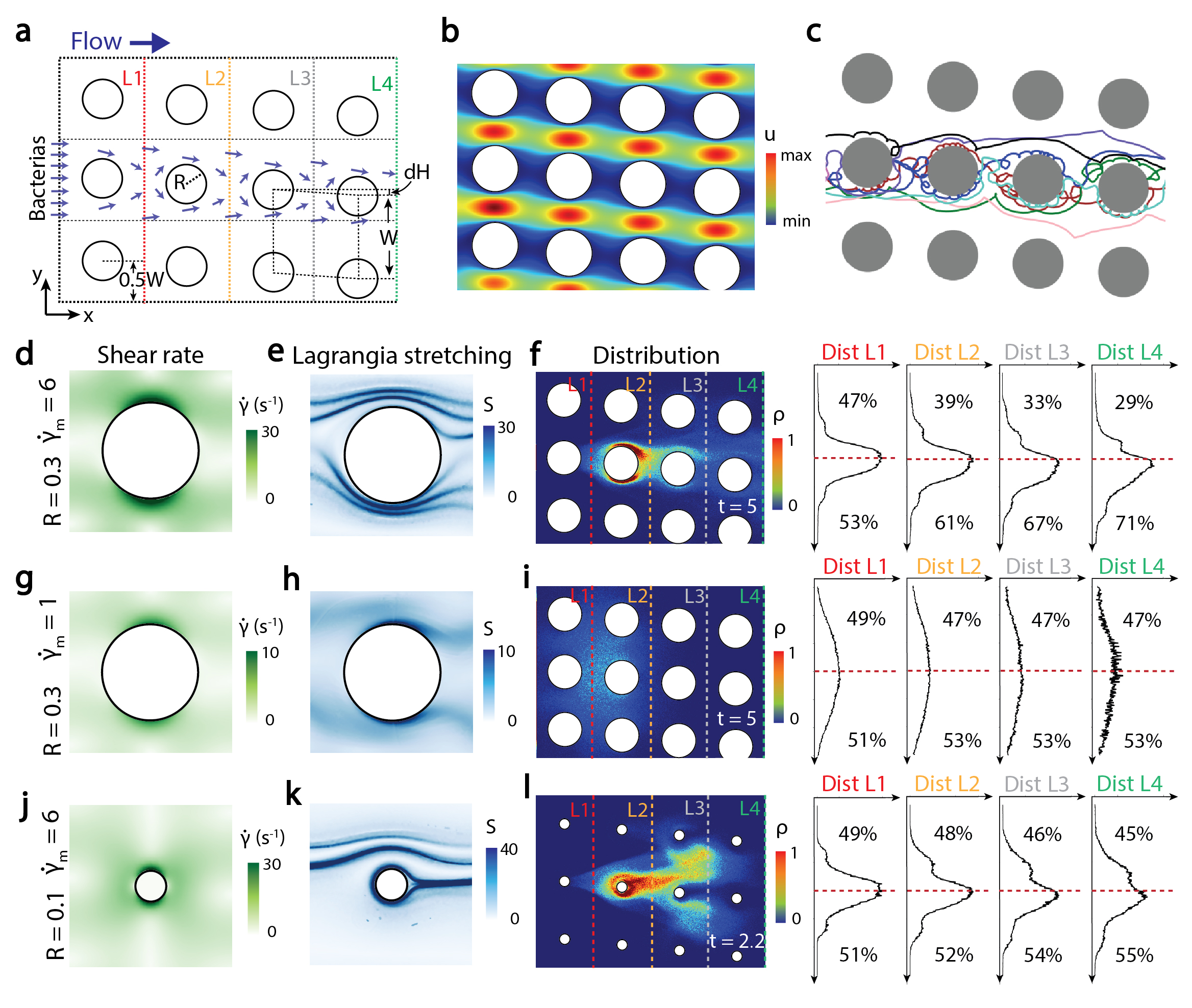}
\end{center}
\setlength{\abovecaptionskip}{-5pt}
\caption{\label{fig:wide}\textbf{The flow field and the microbial distribution in a stretched DLD pillar array.} \textbf{a,} Schematics of a stretched DLD pillar array. The neighboring pillar columns are shifted distance $dH$ in the $y$ direction. \textbf{b,} The distribution of fluidic velocity in the $x$ direction. \textbf{c,} Trajectories of microbes. Colors represent different trajectories. \textbf{d,} The distribution of the shear rate $\dot{\gamma}$, \textbf{e,} the distribution of the Lagrangia stretching, and \textbf{f,} the distribution of microbial density $\rho$ when $R=0.3W$ and $\dot{\gamma}_m=6 \, s^{-1}$. \textbf{g,} The distribution of the shear rate $\dot{\gamma}$, \textbf{h,} the distribution of the Lagrangia stretching, and \textbf{i,} the distribution of microbial density $\rho$ when $R=0.3W$ and $\dot{\gamma}_m=1 \, s^{-1}$. \textbf{j,} The distribution of the shear rate $\dot{\gamma}$, \textbf{k,} the distribution of the Lagrangia stretching, and \textbf{l,} the distribution of the microbial density $\rho$ when $R=0.1W$ and $\dot{\gamma}_m=6 \, s^{-1}$. In \textbf{f}, \textbf{i} and \textbf{l}, left are the 2D distributions at a single time point, and right are the distributions of microbes when they pass line L1, L2, L3 and L4.}
\end{figure*}

We further quantified the effects of pillar radius and fluid mean shear rate on the microbial motility. Specifically, the MSD of a particle can be approximated by the power law as $MSD=D_{\alpha} t^{\alpha}$, where $D_{\alpha}$ is the equivalent diffusion coefficient, and $\alpha$ is the exponent\cite{33}. Our results show that $\alpha$ is close to 2 regardless of the radius $R$ and the mean fluid shear rate $\dot{\gamma}_m$. $D_{\alpha}$ increases over $\dot{\gamma}_m$ and decreases over $R$ (Figure 2a and b). Particularly, a maximum likelihood estimator (MLE) is applied to analyze the advection speeds and dispersion coefficients of microbes. We found that the estimated lateral speed $\hat{V}_x$ and the lateral dispersion coefficient $\hat{D}_x$ increases when fluid flows faster in the pillar array (i.e., when $\dot{\gamma}_m$ increases), but the transverse dispersion coefficient $\hat{D}_y$ decreases over $\dot{\gamma}_m$. This result agrees well with previous findings that hydrodynamic gradients hinder transverse bacterial dispersion in a microfluidic crystal lattice \cite{25}. Moreover, we found that the pillar radius $R$ has a larger influence on the lateral microbial dynamics ($\hat{V}_x$ and $\hat{D}_x$) when $\dot{\gamma}_m$ is large, but influences transverse dynamics $\hat{D}_y$ significantly when $\dot{\gamma}_m$ is small. Large-size pillars hinder both lateral and transverse microbial movements, especially $\hat{V}_x$ and $\hat{D}_y$. When the pillar has an intermediate radius and $\dot{\gamma}_m$ is large, the lateral dispersion coefficient $\hat{D}_x$ is the largest (Figure 2c, d and e).
\begin{figure*}
\begin{center}
\includegraphics[width=2.0\columnwidth]{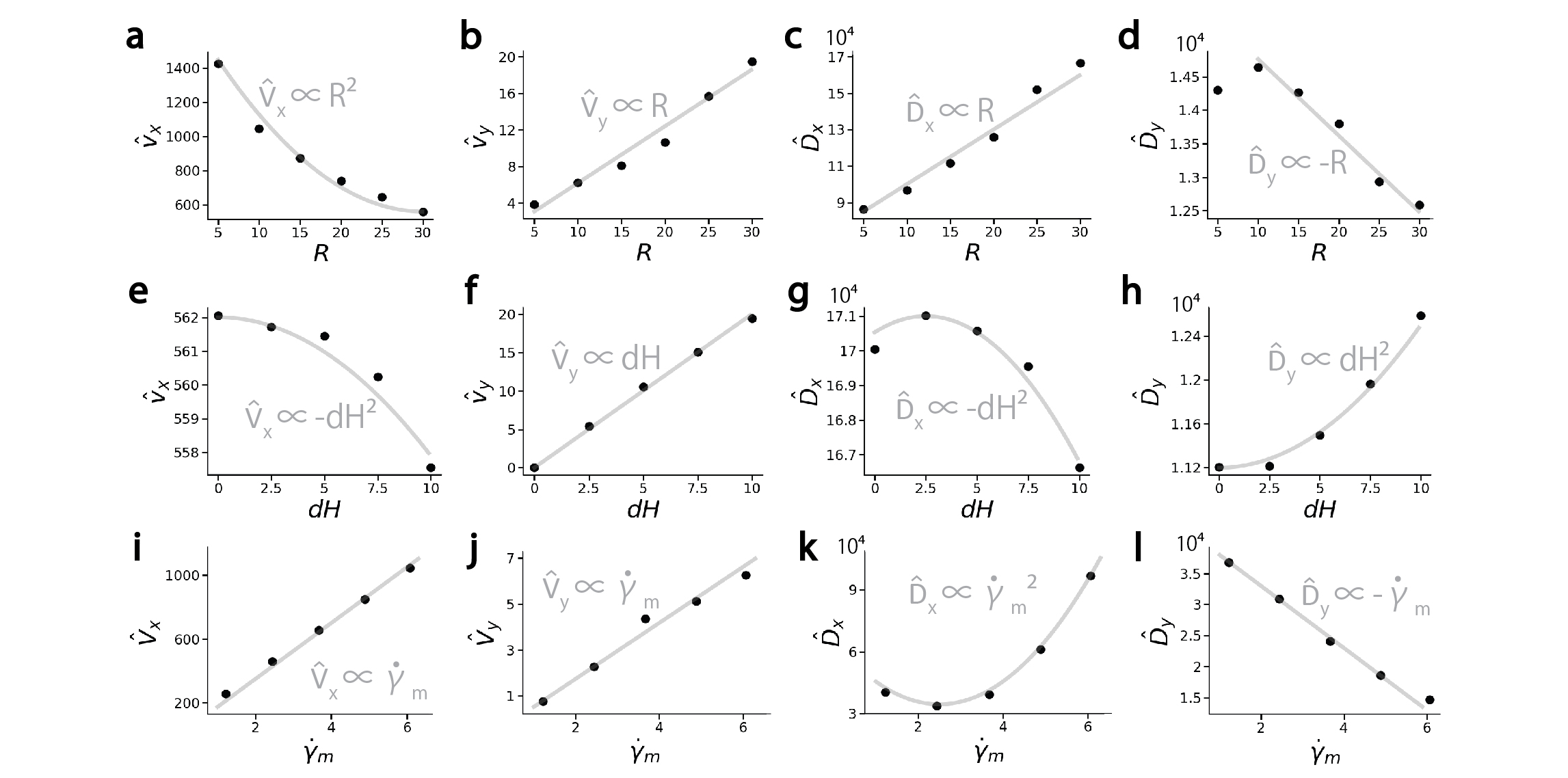}
\end{center}
\setlength{\abovecaptionskip}{-5pt}
\caption{\label{fig:wide}\textbf{The influence of the pillar radius $R$, the asymmetric shift $dH$ and the mean shear rate $\dot{\gamma}_m$ on the lateral and transverse microbial movements.} \textbf{a,} The estimated lateral velocity $\hat{V}_x$ over the pillar radius $R$. \textbf{b,} The estimated transverse velocity $\hat{V}_y$ influenced by $R$. \textbf{c,} The estimated lateral dispersion coefficient $\hat{D}_x$ influenced by $R$. \textbf{d,} The estimated transverse dispersion coefficient $\hat{D}_y$ influenced by $R$. \textbf{e,} The estimated lateral velocity $\hat{V}_x$ over the asymmetric shift $dH$. \textbf{f,} The estimated transverse velocity $\hat{V}_y$ influenced by $dH$. \textbf{g,} The estimated lateral dispersion coefficient $\hat{D}_x$ influenced by $dH$. \textbf{h,} The estimated transverse dispersion coefficient $\hat{D}_y$ influenced by $dH$. \textbf{i,} The estimated lateral velocity $\hat{V}_x$ over the mean shear rate $\dot{\gamma}_m$. \textbf{j,} The estimated transverse velocity  $\hat{V}_y$ influenced by $\dot{\gamma}_m$. \textbf{k,} The estimated lateral dispersion coefficient $\hat{D}_x$ influenced by $\dot{\gamma}_m$. \textbf{l,} The estimated transverse dispersion coefficient $\hat{D}_y$ influenced by $\dot{\gamma}_m$. Black dots are simulation results. Grey lines are fitted curves.}
\end{figure*}

\subsection{The microscopic microbial dynamics surrounding a pillar}
To reveal the mechanisms that cause the transient microbial accumulation and the slow changing of microbial density surrounding the pillar, the microscopic dynamics of microbes is analyzed. The trajectories of microbes in the pillar arrays are first visualized. Figure 3a shows that when the pillar radius $R$ is small, many microbes maintain relatively straight trajectories and flow through the array quickly. Only some trajectories that are close to the pillar are redirected to move around the pillar. However, when the pillar radius $R$ increases, meandering trajectories will dominate. Specifically, three types of movements, ‘swinging’, ‘zigzag’ and ‘adhesive’ movements are observed. When a microbe locates between two rows of pillars, it tends to swing up and down while keeping moving forward. This microbial motion is thus referred to as the ‘swinging’ motion (Figure 3b (1)). When a microbe reaches the vicinity of the pillar, the local velocity gradient applies hydrodynamic torque on the microbe, leading to the frequent loops near the pillar surface (Figure 3a). Though the microbe keeps rotating, it is hardly to swim far away from the top or the bottom sides of the pillar where the shear rate is the largest, as shown in Figure 1f. If the microbe continues swimming from the bottom of the pillar to its leeward side and shifts to the top side of the next pillar, this type of movement is referred to as a ‘zigzag’ mode (Figure 3b (2)). If the microbe moves along the top pillar surface and shifts to the top side of the next pillar, it is then referred to as an ‘adhesive’ mode (Figure 3b (3)). To understand how the pillar radius $R$ and the mean shear rate $\dot{\gamma}_m$ influence the three types of movements, the microbial moving direction $\phi$ in a circular coordinate, where the coordinate center is the pillar center, is calculated. When $\dot{\gamma}_m = 6 \, s^{-1}$, there is a thin dark layer surrounding the pillar surface, where $\phi$ is close to $\pi/2$ (Figure 3c). In this layer, microbes move parallel to the pillar surface, resulting in a ‘zigzag’ or an ‘adhesive’ movement. When $R=0.3W$ and $\dot{\gamma}_m=6 \, s^{-1}$, there is a white area next to the thin dark layer (Figure 3c). In this white area, microbes are moving away from the pillar, which is likely to lead to a ‘swinging’ motion. The white area is hard to be recognized when $R=0.1W$ (Figure 3c). Thus, the ‘swinging’ motion is less frequent when pillar size is reduced. When $\dot{\gamma}_m=1 \, s^{-1}$, the dark thin layer and the white area disappear (Figure 3c), suggesting the microbial trajectories become more randomized. Moreover, the cell density along the red line in Figure 3d in the $y-\theta$ phase space is calculated. When $R=0.3W$ and $\dot{\gamma}_m=6 \, s^{-1}$ (Figure 3e, left), around half of the microbes move toward the pillar leeward side, which is represented by the colorful areas in the left bottom and right upper corners. When the radius R decreases to $0.1W$, the portion of microbes that flow horizontally becomes larger (Figure 3e, right). And when $\dot{\gamma}_m=1 \, s^{-1}$, the two probabilities of a microbe moving toward and away from the pillar get closer (Figure 3e, central). The meandering trajectories and the frequent loops increase the distance a microbe travels, such that the speed of microbial population moving forward is reduced. Figure 3f shows that a larger radius $R$ and a small shear rate $\dot{\gamma}_m$ lead to long moving distances of microbes to reach the areas behind the pillar. Besides, the slower speed of the microbe near the pillar surface (Figure 3a) also affects the speed of the microbial population. Therefore, in the array of large-size pillar, microbial transient accumulation and the slow change of microbial density is observed (Figure S1).

\begin{figure*}
\begin{center}
\includegraphics[width=2.0\columnwidth]{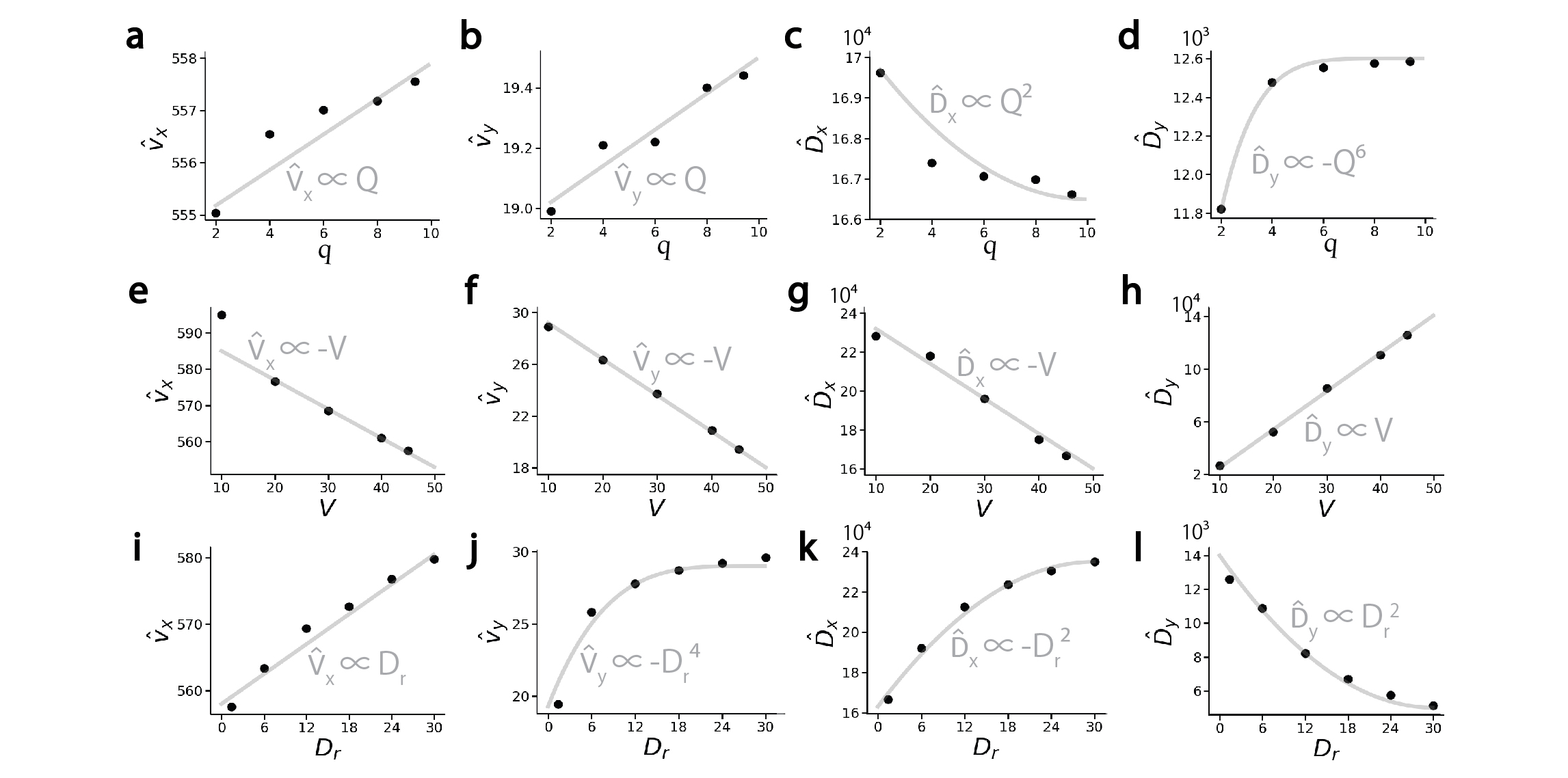}
\end{center}
\setlength{\abovecaptionskip}{-5pt}
\caption{\label{fig:wide}\textbf{The influence of microbial properties, including the aspect ratio $q$, the swimming speed $V$ and the rotational diffusion coefficient $D_r$ on the lateral and transverse microbial movements.} \textbf{a,} The estimated lateral velocity $\hat{V}_x$ over the aspect ratio $q$. \textbf{b,} The estimated transverse velocity $\hat{V}_y$ influenced by $q$. \textbf{c,} The estimated lateral dispersion coefficient $\hat{D}_x$ influenced by $q$. \textbf{d,} The estimated transverse dispersion coefficient $\hat{D}_y$ influenced by $q$. \textbf{e,} The estimated lateral velocity $\hat{V}_x$ over the swimming speed $V$. \textbf{f,} The estimated transverse velocity $\hat{V}_y$ influenced by $V$. \textbf{g,} The estimated lateral dispersion coefficient $\hat{D}_x$ influenced by $V$. \textbf{h,} The estimated transverse dispersion coefficient $\hat{D}_y$ influenced by $V$. \textbf{i,} The estimated lateral velocity $\hat{V}_x$ over the rotational diffusion coefficient $D_r$. \textbf{j,} The estimated transverse velocity $\hat{V}_y$ influenced by $D_r$. \textbf{k,} The estimated lateral dispersion coefficient $\hat{D}_x$ influenced by $D_r$. \textbf{l,} The estimated transverse dispersion coefficient $\hat{D}_y$ influenced by $D_r$. Black dots are simulation results. Grey lines are fitted curves.}
\end{figure*} 

\begin{figure*}
\begin{center}
\includegraphics[width=1.8\columnwidth]{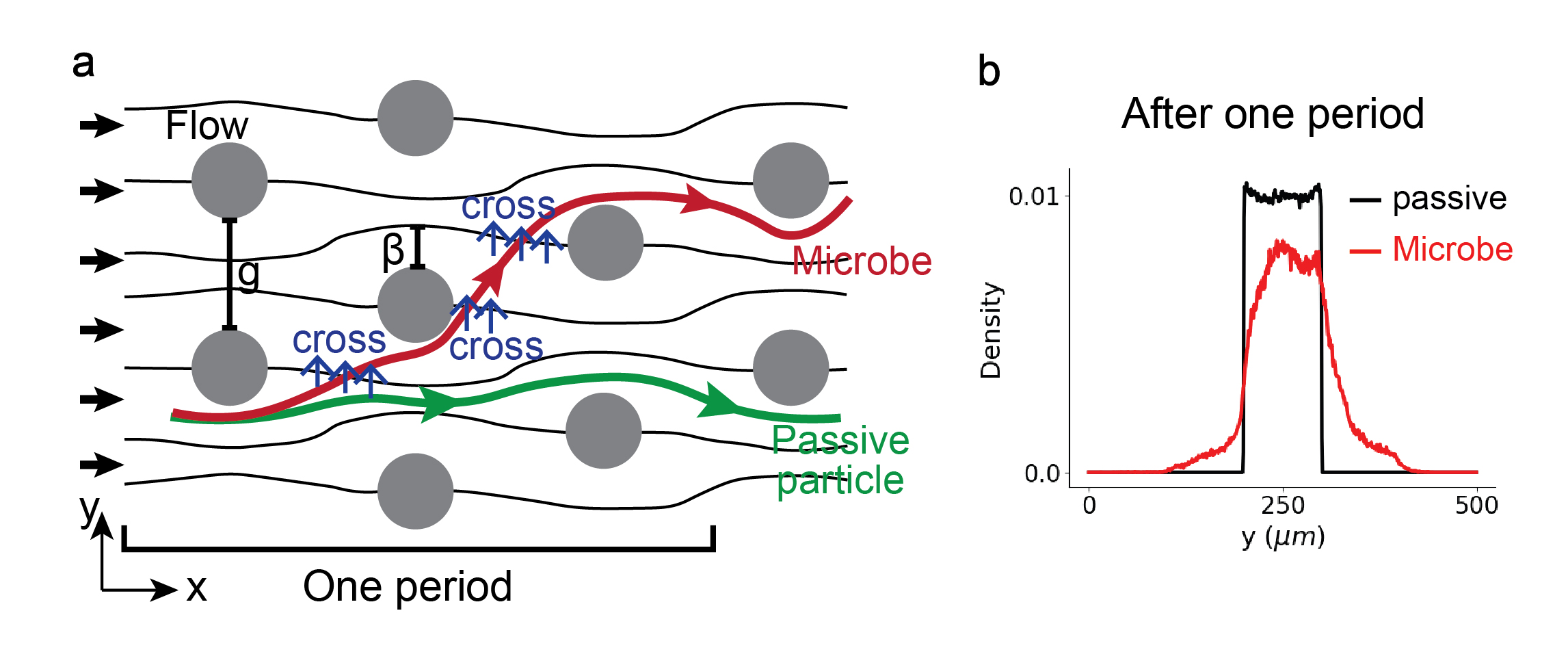}
\end{center}
\setlength{\abovecaptionskip}{-5pt}
\caption{\label{fig:wide}\textbf{Schematics showing the streamlines flowing through a stretched DLD pillar array.} \textbf{a,} Small passive particles are carried unobstructed to a position without any transverse displacements (the green trajectory). Microbes are redirected by the fluid shear and swim across the streamlines, which leads to the transverse migration perpendicular to the flow direction (the red trajectory). \textbf{b,} The distributions of microbes and passive particles along the y direction after passing one period of DLD pillar array (when $x=3W=300\, \mu m$). In the calculation, $W=100 \, \mu m$, $R=30 \, \mu m$, $g=70 \, \mu m$, and the transverse shift $dH=W/3$.}
\end{figure*}

\subsection{Microbes flow through a DLD pillar array}
When the geometric symmetry is broken, few previous experiments \cite{20, 25, 34, 35} has shown that the microbial transport and pattern are influenced by the asymmetric structures. There are two general ways to break the symmetry of a pillar array. One is to rotate the incidence angle of the flow relative to the pillar array, which is also referred to as the rotated square lattice. In this scenario, Dehkharghani et al. \cite{25} found that the microbial densification patterns and the dispersion coefficients are affected by the incidence angle. Another is to shift adjacent columns of pillars by a small distance $dH$ and create a stretched array, which is referred to as a stretched DLD array (or a row-shifted parallelogram array). The stretched DLD array has been used to separate rod-like bacteria \cite{21, 27, 28}. However, since the results in the rotated square lattice layout do not extend to the stretched DLD array, the dynamics and regulatory mechanisms of microbes in the stretched DLD array are still missing.

As shown in Figure 4a, the stretched DLD pillar array contains rows of pillars with radius $R$. Adjacent rows and adjacent columns are separated by a distance $W$. The right neighboring pillar is shifted a distance $dH$ in the $y$ direction. The shift between one column and its $M$ nearest columns is $MdH$, which is chosen to coincide with $W$. Thus, the array is periodic and invariant to a translation $MdH$ in the $y$ direction. And the stretched DLD array has a built-in angle $\theta_{DLD}=\arctan{1/M}$. Fluid flows in the horizontal direction into the pillar array, resulting in a large flow speed between two rows of pillars and a small speed behind each individual pillar (Figure 4b). Two adjacent ellipsoidal areas that have the largest flow velocity $u$ is also shifted $dH$ in the $y$ direction. Microbial trajectories bend around the pillar and are influenced by the built-in angle $\theta_{DLD}$ (Figure 4c). Compared with the symmetric pillar array, the distribution of the shear rate $\dot{\gamma}$ is rotated around the pillar center by $\theta_{DLD}$ (Figure 4d, g and j). The distribution of Lagrangia stretching becomes asymmetric about the $x$ axis (Figure 4e, h and k). When $R=0.3W$ and $\dot{\gamma}_m=6 \, s^{-1}$, the large-stretching regions at the top side of the pillar extend away from the pillar surface, while at the bottom side they concentrate near the pillar surface (Figure 4e). The asymmetric fluid properties induce the macroscopic movements of microbes in the $y$ direction. Figure 4f shows that after passing four columns of pillars, the peak of the microbial density shifts toward the bottom. When $R$ decreases to $0.1W$, the Lagrangia stretching is still large surrounding the pillar surface. But at the top side of the pillar, two streamline-like regions that have large Lagrangia stretching appear. At the bottom side, the Lagrangia stretching remains small (Figure 4k). Consequently, microbes that have the highest density at the center start to separate into two groups. Around 45\% of microbes are in the upper group and move faster than the rest 55\% of microbes in the bottom group. When the mean shear rate $\dot{\gamma}_m$ decreases (i.e. $\dot{\gamma}_m=1\, s^{-1}$), the influence of the asymmetric pillar arrangement on the flow field is small (Figure 4h), and the microbial movements are more randomized (Figure 4i). 

To get a quantitative understanding of the microbial motility in the stretched DLD pillar array, we then estimate the lateral and transverse microbial velocities and dispersion coefficients based on a maximum likelihood estimator (MLE). Results in Figure 5 show that the velocities and the dispersion coefficients are either linearly or parabolically proportional to the parameters that modulate the heterogeneous flow field, including the pillar radius $R$, the asymmetric shift $dH$ and the mean shear rate $\dot{\gamma}_m$. Specifically, when the pillar radius increases, the lateral velocity along the flow direction decreases in a parabolic manner (Figure 5a), while the transverse velocity increases linearly (Figure 5b). The dispersion coefficients have the opposite trend over the pillar radius. The lateral dispersion increases (Figure 5c), but the transverse dispersion is reduced over the radius (Figure 5d). When the asymmetric shift becomes large, the lateral velocity and the dispersion coefficient decreases parabolically (Figure 5e and g), while the transverse motions ($\hat{V}_y$ and  $\hat{D}_y$) are enhanced (Figure 5f and h). As for the mean shear rate, when fluid flows faster ($\dot{\gamma}_m$ increases), the lateral and transverse velocities increase linearly (Figure 5i and j). The lateral dispersion coefficient also increases, but in a parabolic manner (Figure 5k). The transverse dispersion is reduced (Figure 5l). Overall, these results indicate that the structural asymmetry enhances the transverse microbial movements ($\hat{V}_y$ and $\hat{D}_y$). The flow speed (reflected by $\dot{\gamma}_m$) increases microbial advection ($\hat{V}_x$ and $\hat{V}_y$). And large pillars hinder the lateral advection and the transverse diffusion, but improve the transverse advection and the lateral diffusion.

Since the lateral and transverse movements of microbes are the macroscopic reflections of interactions among microbial swimming motion, fluid flow and structural boundaries, the microbial properties should also play significant roles to determine its advection and dispersion. Therefore, we further quantify the effects of three microbial properties, including the aspect ratio $q$, which describes how prolate the microbial body is, microbial swimming speed $V$, and the rotational diffusion coefficient of the microbe $D_r$. Results in Figure 6a and b show that prolate microbes tend to have larger advection speed in both $x$ and $y$ directions. The lateral diffusion coefficient decreases parabolically over the aspect ratio $q$ (Figure 6c). The transverse diffusion coefficient is less influenced when $q$ is larger than 4 (Figure 6d). The swimming speed $V$ has linear effects on the microbial motility. Surprisingly, velocities  $\hat{V}_x$, $\hat{V}_y$ and the lateral diffusion coefficient $\hat{D}_x$ decreases over the swimming speed $V$ (Figure 6e, f and g). Only the transverse diffusion coefficient $\hat{D}_y$ increases over $V$ (Figure 6h). This result suggests that the fast moving of microbial population in the pillar array is mainly attributed to the flow instead of the swimming of microbe individuals. In addition, the lateral microbial movements ($\hat{V}_x$ and $\hat{D}_x$) increase when the rotation (or tumbling) of the microbe is enhanced (Figure 6i and k). The transverse velocity $\hat{V}_y$ is less influenced when the rotational diffusion coefficient $D_r$ is large (Figure 6j). And the transverse dispersion  $\hat{D}_y$ decreases over $D_r$ (Figure 6l). It should also be noted that the influence of the aspect ratio $q$ is much smaller than the influence of the swimming speed $V$ and the rotational diffusion coefficient $D_r$ (Figure 6). In summary, prolate microbes that swim slowly but can tumble quickly are likely to have large advection velocities and also large lateral dispersion in the stretched DLD pillar arrays. In contrast, prolate microbes that swim fast and tumble less frequently usually have large transverse dispersion but small velocities and small lateral dispersion.

\section{Discussion}
In this work, we quantitatively studied the microbial dynamics influenced by the heterogeneous flow field in both symmetric and DLD pillar arrays based on a Langevin model. Due to the interplay among swimming microbes, the flowing fluid and pillars, microbes move with a slower speed and meandering trajectories near the pillar surfaces, which leads to the transient accumulation of a large portion of microbes and the sluggish microbial transport in the vicinity of pillars. When the pillar size is large and the fluid flows quickly, large fluid shear causes the swinging, zigzag and adhesive motions of microbes. Particularly in an asymmetric DLD pillar array, microscopic microbial movements induce the deterministic shift of the microbial population perpendicularly to the flow direction. Moreover, a comprehensive analysis of the effects of the pillar size and arrangement, fluid properties and microbial properties on the lateral and transverse microbial advection and dispersion is provided.

Our model and studies provide useful insights for designing stretched DLD pillar arrays for microbial separation and harvesting. The stretched DLD pillar array (Figure 4a) has long been used to separate micro/nano sized particles when the Péclet number is large and the particle diffusion is negligible \cite{20, 21, 22, 23, 24, 26, 28, 36}. The underlying separation mechanism is that small passively-advected particles are carried unobstructed to a position without any transverse displacements. But particles larger than the critical diameter $d_c$ (i.e., $d_c=2\beta=2g/3$ in Figure 7) would bump with the pillars and lead to trajectories that follow the built-in angle $\theta_{DLD}$ \cite{20, 22}. Recently, Kim et al. \cite{27} improved the theory and showed that the pillar array geometry distorts the flow such that particle trajectories have migration angles between 0 and  $\theta_{DLD}$. Here, we firstly considered the active swimming of microbes in the stretched DLD pillar array and showed that even sufficiently small microbes can manifest diverse movement patterns and meandering trajectories due to their active motions. Specifically, when the microbes enter the pillar array with a uniform distribution at $200 \, \mu m \leq y \leq 300 \, \mu m$, the fluid shear redirects microbes such that many microbes swim across the streamlines (Figure 7a). Intuitively, when a microbe has a larger swimming speed than the local fluidic speed, it is more likely to swim against the flow and cross the streamlines. In a macroscopic point of view, after passing one period of pillar arrays (i.e. three columns of pillars in Figure 7a), a considerable fraction of microbes has a transverse migration and the migration angle is between 0 and $\theta_{DLD}$ (the red curve in Figure 7b). Without the active swimming motion, passive particles maintain the same uniform distribution after passing the three pillar columns (the black curve in Figure 7b). The transverse migration of microbial population is affected by the pillar layout and the fluidic properties (Figure 5\&6). Thus, our model and results provide quantitative references to the design of DLD pillar arrays to separate and harvest microbes that have similar sizes but different swimming abilities. Particularly, our results indicate that a pillar array with a large pillar radius $R$, a small gap between two pillars $g$ and a large asymmetry shift $dH$ has a great capability to separate microbes that have distinct swimming speeds as well as to separate the mixture of microbes and other passive particles such as proteins.

In the natural habitats of microbes, different types of environmental fluctuations such as nutrients and sunlight occur in various timescales, from seconds to seasons, and have a large spatial heterogeneity. For example, in the sandy segments, oxygen concentrations fluctuate on minute scales due to the changing currents and sediment movements \cite{6}. In the soil pores, plant roots release labile exudates and create short pulses of microscale nutrients in the surrounding areas \cite{37, 38}. Thus, not only the spatial microbial distribution, which has been investigated intensively in recent years \cite{29, 39, 40}, but also the temporal microbial evolution \cite{41, 42} play important roles in microbial behaviors and ecology. In the pillar array which can be regarded as a simple and ideal model of the porous media, we observed the sluggish transport and transient accumulation of a large amount of microbes in the vicinity of pillars. This temporal characteristic of microbial movements exists widely as long as a curving solid surface induces proper local fluid shear which leads to looping trajectories and slow moving speed of microbes. So, it is likely that the sluggish microbial transport would coordinate with the temporal fluctuations in the environment and affects microbial growth, proliferation and community formation (i.e., formation of biofilms), which could be an important and interesting future direction.

Overall, the quantitative mechanistic model and findings we have proposed in this study open new frontiers in the possibility of controlling the spatial and temporal evolution of microbial population, which can be greatly beneficial to many ecological and technological applications, such as water filtration, bioremediation and carbon fixation.

\section*{Supplementary material}
See supplementary material for the spatial evolution of microbial distributions influenced by the pillar radius $R$ in the symmetric pillar array.

\begin{acknowledgments}
We acknowledge support from the National Natural Science Foundation of China (No. 12102081) and Fundamental Research Funds for the Central Universities in China (DUT21RC(3)044).
\end{acknowledgments}

\section*{References}
\nocite{*}
\bibliography{manuscript_bib}

\end{document}